# Charge compensation and optimal stoichiometry in superconducting $(Ca_xLa_{1-x})(Ba_{1.75-x}La_{0.25+x})Cu_3O_y$

Dale R. Harshman[1,2,*] and Anthony T. Fiory[3]

[1] *Physikon Research Corporation, Lynden, Washington 98264, USA*
[2] *Department of Physics, University of Notre Dame, Notre Dame, Indiana 46556, USA*
[3] *Department of Physics, New Jersey Institute of Technology, Newark, New Jersey 07102, USA*

The superconductive and magnetic properties of charge-compensated $(Ca_xLa_{1-x})(Ba_{1.75-x}La_{0.25+x})Cu_3O_y$ (normally denoted as CLBLCO) are considered through quantitative examination of data for electrical resistivity, magnetic susceptibility, transition width, muon-spin rotation, x-ray absorption, and crystal structure. A derivative of $LaBa_2Cu_3O_y$, cation doping of this unique tetragonal cuprate is constrained by compensating La substitution for Ba with Ca substitution for La, where for $0 \leq x \leq 0.5$ local maxima in $T_C$ occur for $y$ near 7.15. It is shown that optimum superconductivity occurs for $0.4 \leq x \leq 0.5$, that the superconductivity and magnetism observed are nonsymbiotic phenomena, and that charge-compensated doping leaves the carrier density in the cuprate planes nearly invariant with $x$, implying that only a small fraction of superconducting condensate resides therein. Applying a model of electronic interactions between physically separated charges in adjacent layers, the mean in-plane spacing between interacting charges, $\ell = 7.1206$ Å, and the distance between interacting layers, $\zeta = 2.1297$ Å, are determined for $x = 0.45$. The theoretical optimal $T_{C0} \propto \ell^{-1}\zeta^{-1}$ of 82.3 K is in excellent agreement with experiment ($\approx 80.5$ K), bringing the number of compounds for which $T_{C0}$ is accurately predicted to 37 from six different superconductor families (overall accuracy of $\pm 1.35$ K).

## I. INTRODUCTION

The charge-compensated compound $(Ca_xLa_{1-x})(Ba_{c-x}La_{2.0-c+x})Cu_3O_y$ with $c = 1.75$ is a derivative of $LaBa_2Cu_3O_{7-\delta}$ wherein $(0.25 + x)$ $La^{+3}$ substituting for $Ba^{+2}$, coupled with $(x)$ $Ca^{+2}$ substituting for the $La^{+3}$ separating the cuprate planes, preserves a constant cation charge value (exclusive of the Cu ions) of $Q = +7.25$,[1] hence the "charge-compensated" denotation. Early experiments exploring the *c-x-y* formulation matrix determined a maximum transition temperature of $T_C^{max} = 82$ K by studying the dependence of $y$ relative to the metal-insulator transition value $y_{M-I}$.[2] Subsequent investigations provided more detail on the $y$ dependence of $T_C$, and nearly the same $T_C^{max} = 81$ K occurs for $x = 0.4$, and $y$ in the vicinity of 7.15.[3,4] The authors of Ref. 2 obtained a unique experimental result that corresponds to the optimum $T_{C0}$, as defined in Ref. 5 and herein. The later researchers[3,4,6] introduce a different scaling phenomenology: $y$ is offset by a model function of $x$ and $T_C$ is scaled to a redefined $T_C^{max}$ that depends on $x$ (this redefinition is discussed further in Sec. IV).

For $(Ca_xLa_{1-x})(Ba_{1.75-x}La_{0.25+x})Cu_3O_y$, the optimal superconductive material is most closely realized by the stoichiometry corresponding to $T_C^{max}$ (as originally defined[2]). Deviations from optimum stoichiometry (yielding $T_C < T_{C0}$) typically introduce inhomogeneities which degrade the superconductive quality of the material, manifested as depressed resistive superconducting transitions (considered generally in Refs. 7 and 8). For example, maintaining a constant $Q$ by controlling the average cation valence does not prevent changes in the charge available for superconductivity, nor the degradation of the superconducting condensate caused by widely varying $c$, $x$ or $y$ from their optimum values. Consequently, $(Ca_xLa_{1-x})(Ba_{1.75-x}La_{0.25+x})Cu_3O_y$ materials grown with $x < 0.4$, as shown by the data presented in Sec. II, are nonoptimal.

Owing to the charge-compensated nature of $(Ca_xLa_{1-x})(Ba_{1.75-x}La_{0.25+x})Cu_3O_y$, it is instructive to consider the optimal material (achieved for $0.4 \leq x \leq 0.5$) from the perspective of a new theoretical treatment that assumes the pairing is mediated via Coulomb interactions between physically separated carrier bands;[5] $YBa_2Cu_3O_{7-\delta}$ (90-K phase optimized for $\delta = 0.08$) and other optimal high-$T_C$ compounds have been found to contain both holes and electrons confined in two dimensions and separated physically in different regions of the unit cell.[9,10] From this, and the fact that phonon (or polaron) mediation has been ruled out,[11] it is natural to expect that Coulomb forces between the holes and electrons would dominate the superconductive pairing. At the time of this writing, this model has been validated (with a statistical error between the calculated and measured $T_{C0}$ of $\pm 1.34$ K) for 36 different materials from six superconducting families (cuprates, ruthenates, rutheno-cuptates, iron pnictides, iron chalcogenides, and organics), with measured $T_{C0}$ values ranging from 10 K to 150 K.[12,13] The present application of this model to $(Ca_xLa_{1-x})(Ba_{1.75-x}La_{0.25+x})Cu_3O_y$ gives $T_{C0} = 82.3$ K (see Sec. III).

Section II presents an analysis of prior experimental data to help clarify the unique experimental result identifying $T_C^{max}$ for $(Ca_xLa_{1-x})(Ba_{1.75-x}La_{0.25+x})Cu_3O_y$, showing that the optimal charge-compensation variable $x$ lies in the range 0.4 to 0.5. Relevant observations include the nonproportional variation between nonoptimum $T_C$ and muon-spin depolarization rate, nuclear quadrupole resonance (NQR) frequencies, and x-ray absorption spectroscopy (XAS) measurements of cuprate-plane charges showing a near invariance with $x$. In Sec. III, the theory for $T_{C0}$ derived from pairing by the interlayer Coulomb interaction between physically separated charge layers is briefly described, and $T_{C0}$ for optimal $(Ca_xLa_{1-x})(Ba_{1.75-x}La_{0.25+x})Cu_3O_y$ is calculated. Section IV provides discussions of oxygen stoichiometry, superconducting inhomogeneity and locus, magnetic behavior, and the measurements-based rationale for distinguishing between optimum and non-optimum materials. Conclusions are drawn in Sec. V.

## II. SUMMARY OF EXPERIMENTAL RESULTS

Early electrical resistivity and magnetic susceptibility data on $(Ca_xLa_{1-x})(Ba_{1.75-x}La_{0.25+x})Cu_3O_y$, measured as functions of $x$ and $y$, clearly exhibit nonoptimal behavior for $x < 0.4$ and for any value of $y$. The experimental evidence includes transition temperatures $T_C^R$ and $T_C^\chi$, and associated transition widths $\Delta T_C^R$ and $\Delta T_C^\chi$, from resistivity and susceptibility, respectively, for values of $x$ ranging from 0 to 0.50 and $y \approx 7.1$.[1] Later results were obtained from superconducting onset constructions that typically yield higher transition points.[2,4] The various measurements of the highest (or nearly the highest) transition temperature at each $x$ are plotted as functions of $x$ in Fig. 1, illustrating the trend: for $x \sim 0.45$ to 0.50, $T_C$ is maximum and approaches the $T_C^{max}$ of 82 K reported in Ref. 2 and 81 K reported in Refs. 3 and 4. Also marked in Fig. 1 is the $T_{C0}$ calculated in Sec. III B, which nearly coincides with the observed maximum $T_C$.

The inset in Fig. 1 shows that there is a finite pressure dependence of $T_C$ ($dT_C/dP > 0$),[2] which introduces some uncertainty in establishing an upper bound for $T_C^{max}$. That $dT_C/dP$ turns sharply downward as $T_C$ approaches $T_C^{max}$ suggests that any pressure correction on the extrapolated $T_C^{max}$ is likely to be small.

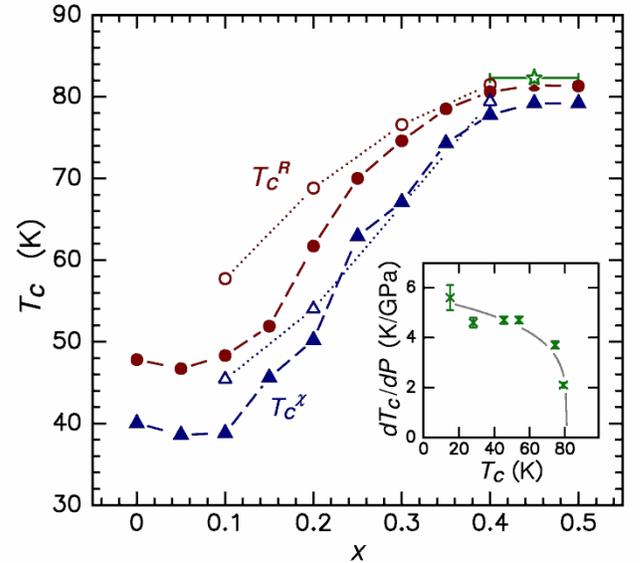

FIG. 1. Transition temperatures $T_C$ for $(Ca_xLa_{1-x})(Ba_{1.75-x}La_{0.25+x})Cu_3O_y$, maximized by oxygen content $y$, extracted from electrical resistivity (filled circles after Ref. 1, open circles after Ref. 4) and magnetic susceptibility (filled triangles after Ref. 1, open triangles after Ref. 2) as a function of $x$. Star symbol denotes theoretical $T_{C0}$ averaged over $x = 0.4 - 0.5$. Inset represents $dT_C/dP$ vs $T_C$ after Ref. 2; curve denotes trend.

Specific experimental indicators of materials quality are extractable from temperature-dependent resistivity $\rho(T)$. By fitting a line to $\rho(T)$ just above $T_C$ (using data plotted in Ref. 1 for analysis), the extrapolated normal-state resistivity at $T_C$, denoted as $\rho(T_C^+)$, and the extrapolated zero-temperature residual resistivity, denoted as $\rho(0)$, have been determined in this work. These resistivity data together with the associated percentage transition widths $\Delta T_C/T_C$ are plotted in Fig. 2. Viewed as functions of $x$, it is clear that the extrapolated $\rho(0)$ trends from large values at low $x$ to very small values in the vicinity of $x \sim 0.45$ to 0.50, signaling that $T_C^{max}$ corresponds to minimization of electronic scattering and disorder. The similar downward trend of the normal-state resistivity, represented by $\rho(T_C^+)$, to values under 0.5 m$\Omega$ cm indicates a corresponding approach to good metal transport. The associated decreasing trend in $\Delta T_C/T_C$ shows that samples for $x \geq 0.4$ have the most uniform superconducting states, which is consistent with the indicators of lowest disorder from normal-state transport.

Superconductivity in $(Ca_xLa_{1-x})(Ba_{1.75-x}La_{0.25+x})Cu_3O_y$ has also been qualified for various $x$ and $y$ by means of positive muon-spin-rotation spectroscopy ($\mu^+$SR), where a vortex state is induced in an applied magnetic field and the time $t$ dependence of muon-spin polarization is fitted with a Gaussian model, $\exp(-\sigma_\mu^2 t^2/2)$. Results for $\sigma_\mu/\gamma_\mu$, where $\gamma_\mu = 8.514 \times 10^5$ $G^{-1}s^{-1}$ is the muon gyromagnetic ratio, provide estimates of the superconducting magnetic penetration depth $\lambda$ through the theoretical relation, $\sigma_\mu/\gamma_\mu = (1.26 \times 10^{-8}$ G cm$^2) \lambda^{-2}$, where for ceramic specimens of $(Ca_xLa_{1-x})(Ba_{1.75-x}La_{0.25+x})Cu_3O_y$, as well as $YBa_2Cu_3O_{7-\delta}$, $\lambda^{-2}$ is obtained as a crystallographic average dominated by the basal ($a$-$b$) plane component.[14-16] Under optimal experimental and materials conditions, which require a nearly perfectly formed and static lattice of fluxons within a homogeneous clean superconducting material,[17] the zero-temperature limit, denoted as $\lambda(0)$, yields the London penetration depth $\lambda_L = (m^*c^2/4\pi n_s e^2)^{1/2}$, reflecting the intrinsic superconducting carrier density $n_S$ and an effective mass $m^*$.[18] On the other hand, the presence of disorder and inhomogeneity introduces residual normal conductivity that increases the measured $\lambda(0)$ relative to the intrinsic penetration depth (discussed, e.g., in Ref. 19).

In superconducting material of high quality, the residual mean-free path $\ell_{mfp}(0)$ is generally larger than the Pippard coherence distance $\xi_0$, obeying a criterion that is equivalent to $\hbar\tau^{-1}(0)/k_BT_C < 1$, where $\tau^{-1}(0)$ is the residual scattering rate ($\hbar$ is reduced Planck's constant; $k_B$ is the Boltzmann constant). It is useful to estimate $\tau^{-1}(0)$ from $\lambda(0)$ and the extrapolated residual resistivity $\rho(0) = m^*\tau^{-1}(0)/ne^2$, where $n$ is the normal-state carrier density, by equating the normal $m^*/n$ to the superconducting $m^*/n_S$. The result, $\tau^{-1}(0) = 4\pi c^{-2}\lambda^{-2}(0)\rho(0)$, has been evaluated using measurements of $\sigma_\mu$ from Ref. 6. The ratio $\hbar\tau^{-1}(0)/k_BT_C$ is plotted in the inset of Fig. 2, where the horizontal dashed line denotes unity. Points falling above the dashed line indicate low quality for specimens with $x < 0.4$. One notes that this measure of quality is obtained more accurately at low $x$, where $\rho(0)$ becomes comparable to $\rho(T_C^+)$ [see Fig. 2]; moreover, $\lambda^{-2}(0)$ determined from $\sigma_\mu$ is an upper bound at low $x$, since it is uncorrected for disorder.[19]

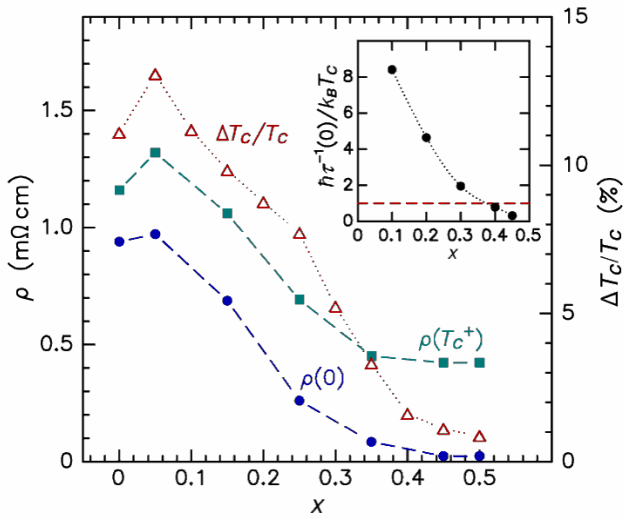

FIG. 2. Resistivity measurements $\rho(T_C^+)$ and $\rho(0)$, left scale, and the percentage transition widths $\Delta T_C/T_C$, right scale, plotted as functions of $x$ for $(Ca_xLa_{1-x})(Ba_{1.75-x}La_{0.25+x})Cu_3O_y$, with $y$ set to maximize $T_C$ (after Ref. 1). Inset shows $\hbar\tau^{-1}(0)/k_BT_C$ plotted against doping $x$; horizontal dashed line denotes unity.

Figures 1 and 2, which represent data corresponding to the highest $T_C$ for a given $x$, are consistent with identifying a unique $T_C^{max}$ for the optimum transition temperature of $(Ca_xLa_{1-x})(Ba_{1.75-x}La_{0.25+x})Cu_3O_y$, validating the conclusions drawn in Ref. 2, which were deduced by scaling methodologies involving $x$, $y$, and $c$. Alternative interpretations have been derived from variations in superconducting and magnetic properties with $x$-$y$ stoichiometry[20-22] and are discussed in Sec. IV.

Values of $T_C$ and $\sigma_\mu$ (low-temperature limit) from $\mu^+$SR measurements of $(Ca_xLa_{1-x})(Ba_{1.75-x}La_{0.25+x})Cu_3O_y$ samples with various $x$ and $y$ were previously presented in Ref. 6, including some sketches to suggest curved or looped trends. The data corresponding to $x = 0.4$ are represented in Fig. 3 as filled circles (underdoped with respect to optimum $y$) and filled triangles (overdoped with respect to optimum $y$). Data corresponding to the highest $T_C$ for $x = 0.1$ and $x = 0.4$ are represented by the filled squares and distinguished by the arrows. As noted in Ref. 6, variations of $T_C$ with $\sigma_\mu$ tend to be insensitive to whether materials are underdoped or overdoped. In particular, the datum for $x = 0.1$ is observed to commingle with data for non-optimal samples, e.g., the filled square is adjacent to a filled circle (data for $x = 0.2$ and $x = 0.3$ in Ref. 6 display the same type of behavior).

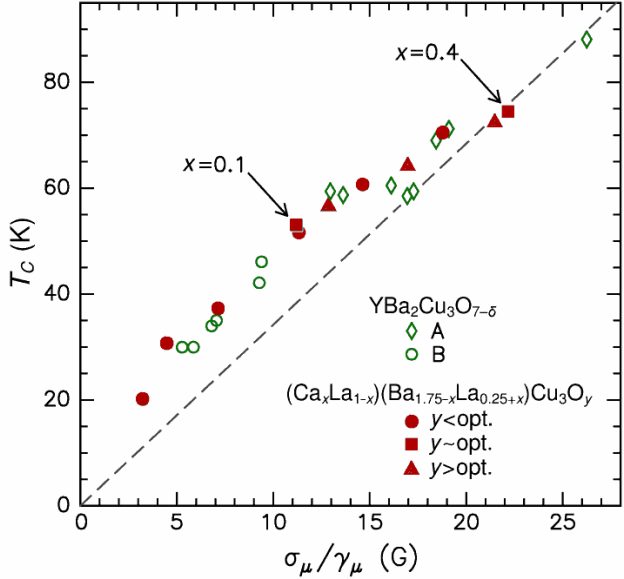

FIG. 3. Transition temperature measured by $\mu^+$SR plotted against the $\mu^+$SR Gaussian relaxation parameter divided by the muon gyromagnetic ratio, $\sigma_\mu/\gamma_\mu$, for polycrystalline samples of $(Ca_xLa_{1-x})(Ba_{1.75-x}La_{0.25+x})Cu_3O_y$ with depleted oxygen at $x = 0.4$ (filled circles), excess oxygen at $x = 0.4$ (filled triangles), and optimum oxygen at $x = 0.1$ and $x = 0.4$ (filled squares, distinguished by arrows) after Ref. 6. Similarly obtained $\mu^+$SR data for ceramic samples of $YBa_2Cu_3O_{7-\delta}$ for various $\delta$ are presented for comparison; A (open diamonds) after Ref. 23 and B (open circles) after Ref. 24. Dashed line is reproduced from Ref. 26.

For comparison, Fig. 3 includes similarly obtained $\mu^+$SR data for $YBa_2Cu_3O_{7-\delta}$ (open diamonds represent data set A from Ref. 23; open circles represent data set B from Ref. 24). For $YBa_2Cu_3O_{7-\delta}$ the variation of $T_C$ in this case is induced solely by oxygen depletion, which has the notable propensity of creating percolative superconductivity (e.g., as observed in recent heat capacity studies[25]). The dashed line in Fig. 3, which has been used to reference $\mu^+$SR data in several previous publications (e.g., Refs. 6 and 26–28), is discussed in Sec. IV. In the case of $YBa_2Cu_3O_{7-\delta}$, many of the data points fall to the left of the dashed line, which represent depressed $\sigma_\mu$ for a given $T_C$; this corresponds to $\lambda > \lambda_L$ and is consistent with non-optimal superconducting samples. The notable exception is the ortho-II phase $YBa_2Cu_3O_{6.60}$ containing ordered chain oxygens; the point for this optimum 60-K superconductor is near the dashed line. It is quite evident that much of the data for $(Ca_xLa_{1-x})(Ba_{1.75-x}La_{0.25+x})Cu_3O_y$ display a pattern of overlapping the results for these nonoptimal $YBa_2Cu_3O_{7-\delta}$ materials, especially in the region $T_C < 60$ K. The possible exception is the $x = 0.4$ sample with highest $T_C$ and $\sigma_\mu$, and indicated by a filled square in Fig. 3; for this sample, both $x$ and $y$ appear to be closest to optimum. This $\mu^+$SR result is also consistent with evidence from resistivity and susceptibility that optimal $(Ca_xLa_{1-x})(Ba_{1.75-x}La_{0.25+x})Cu_3O_y$ corresponds to $x$ of at least 0.4.

Other important results have been reported,[29] where integrated XAS of the $Cu_{2p}$ $L_3$ edge and $O_{1s}$ $K$ edge of $(Ca_xLa_{1-x})(Ba_{1.75-x}La_{0.25+x})Cu_3O_y$ each reveal very weak variation with $x$ at optimum $y$ (less than one standard deviation and about two standard deviations, respectively, for a 0.4-change in $x$), interpreted in Ref. 29 as near invariance of the charge per planar Cu and O with $x$. Conversely, $Cu_{2p}$ $L_3$ and $O_{1s}$ $K$ XAS do detect the expected change in charge on both Cu and O with variation of $y$ at constant $x$, confirming that XAS is indeed sensitive to carrier density variations associated with the cuprate-planes. The charge invariance of the planar Cu and O ions with $x$ provides an independent rationale for grouping together the varied stoichiometries of $(Ca_xLa_{1-x})(Ba_{1.75-x}La_{0.25+x})Cu_3O_y$, such as the various $x$ at local optimum $y$ presented in Figs. 1 and 2.

Independence of the carrier density in the cuprate planes for variable $x$ and constant $y$ may be understood in terms of charge compensation doping in $(Ca_xLa_{1-x})(Ba_{1.75-x}La_{0.25+x})Cu_3O_y$. Consider the layers adjacent to the cuprate planes from the perspective of charge balance. The cation charge state in the $Ba_{1.75-x}La_{0.25+x}O_2$ layers increases as $x/2$ per layer; the cation charge state in the $Ca_xLa_{1-x}$ layer decreases by $x$, which is a decrease of $x/2$ per cuprate plane; hence, the sum of the charges on the cations in the two layers adjacent to a given cuprate plane remains constant with $x$.

### III. THEORETICAL FORMULATION

The theoretical model of Ref. 5 of the pairing mechanism governing high-$T_C$ superconductivity assumes that the pairing interaction occurs between physically separated carrier bands via Coulomb interactions. In the case of $YBa_2Cu_3O_{7-\delta}$, for example, the interacting charges are located in adjacent BaO and $CuO_2$ layers, separated perpendicularly by an interaction distance $\zeta$, with the former designated as part of the type I reservoir (i.e., BaO-CuO-BaO) and the latter assigned to the type II reservoir (i.e., $CuO_2$-Y-$CuO_2$). Here, the CuO and Y layers are referred to as central doping (or inner) layers and the BaO and $CuO_2$ layers as interacting (or outer) layers. Defining $\ell$ to be the mean in-plane spacing between interacting charges, the optimal transition temperature $T_{C0}$, corresponding to the highest $T_C$ for a given compound structure and doping, is given by the algebraic equation,

$$T_{C0} = k_B^{-1} \beta / \ell\zeta = k_B^{-1} \beta (\sigma\eta/A)^{1/2} / \zeta, \qquad (1)$$

where $\eta$ is the number of charge-carrying layers in the type II reservoir (e.g., $\eta = 2$ for $YBa_2Cu_3O_{7-\delta}$ and $(Ca_xLa_{1-x})(Ba_{1.75-x}La_{0.25+x})Cu_3O_y$, corresponding to the two $CuO_2$ planes), $\sigma$ is the fractional charge per outer type I layer per formula unit for participating carriers, $A$ is the basal-plane area per formula unit, $\ell = (\sigma\eta/A)^{-1/2}$, and $\beta$ (= 0.1075 ± 0.0003 eV Å$^2$) is a universal constant, where $\beta/e^2$ has dimensions of length (0.00747 Å) and is equal to about twice the reduced electron Compton wavelength. The optimization of the superconducting state is achieved when the interacting carrier densities associated with the type I and type II reservoirs are in equilibrium, with superconductivity in these systems occurring for $\zeta/\ell \leq 1$.[5]

#### A. Application of Eq. (1)

In general, doping may be either cation or anion and can occur in either reservoir[5] or in both. For compounds (such as $La_{1.837}Sr_{0.163}CuO_4$) where the doping is clearly known, $\sigma$ can be determined directly by allocating the doping charge to the structural layers and charge reservoirs through application of (a first set of) rules (1a) and (1b) that are discussed in Refs. 5 and 12. However, for those compounds in which the doping charge cannot be discerned independently, e.g., $BiSr_2CaCu_2O_{8+\delta}$, $\sigma$ can be calculated by scaling to $\sigma_0$ (= 0.228) for $YBa_2Cu_3O_{6.92}$ according to the formula,

$$\sigma = \gamma \sigma_0, \qquad (2)$$

where $\gamma$ is the product of individual scaling factors $\gamma_i$, which are all dependent upon structure and/or charge state. Since one is primarily interested in determining the charge in the type I interacting layers, $\gamma$ is by default equal to $\gamma_I$ (i.e., the scaling factor corresponding to the type I reservoir). The following (second set of) valency scaling rules are used to determine the $\gamma_i$ comprising $\gamma$ in Eq. (2).

(2a) Heterovalent substitution [in the type I central layer(s)] of a valence +3 ion mapped to a valence +2 ion corresponding to the $YBa_2Cu_3O_{7-\delta}$ structural type introduces a factor of 1/2 in $\gamma$.

(2b) The factor $\gamma$ scales with the +2 (–2) cation (anion) structural and charge stoichiometry associated with participating charge.

(2c) The factor $\gamma$ scales with the net valence of the undoped mediating layer.

Note that the first two valence scaling rules above are defined to be applicable to the type I reservoirs of the cuprate-plane containing superconductors (doping is typically introduced by cation substitution in the type I reservoirs).[30] For example, rule (2b) requires scaling the +2 ion content (as applied to the outer layers) or the number of layers [applied, e.g., to the central donor layer(s) of $Bi_2Sr_2CaCu_2O_{8+\delta}$ (Ref. 5)] to that of $YBa_2Cu_3O_{6.92}$. For the particular case of $(Ca_xLa_{1-x})(Ba_{1.75-x}La_{0.25+x})Cu_3O_y$, only rule (2b) is used with specific regard to the relative $Ba^{+2}$ content.

#### B. Calculation of $T_{C0}$

From Fig. 1, the maximum transition temperature for $(Ca_xLa_{1-x})(Ba_{1.75-x}La_{0.25+x})Cu_3O_y$ occurs for $0.4 \leq x \leq 0.5$, with a value ranging between 77.8 – 79.2 K from (onset) susceptibility, and 79.3 – 80.6 K from resistivity, obtained by taking onset values minus the transition widths; samples for these data correspond to $y = 7.135 – 7.141$.[1] For the mid-range point, $x = 0.45$ and $y = 7.135$, the resistive zero occurs at $T_C^{max} = 80.5$ K, which for the purposes of this paper will be taken to be the experimentally measured $T_{C0}$ (this stoichiometry also produces the sharpest magnetic transition[1]).

Determining σ, corresponding to the optimal compound of this series, is accomplished by utilizing the valence scaling rule (2b) for the type I outer layers. Application of this rule requires scaling the outer layer $Ba^{+2}$ content of $1.75 - x$ in $(Ca_xLa_{1-x})(Ba_{1.75-x}La_{0.25+x})Cu_3O_y$ with respect to the two $Ba^{+2}$ ions in stoichiometric $YBa_2Cu_3O_{6.92}$. Thus γ is given by $(1.75 - x)/2$, such that one then has from Eq. (2),

$$\sigma = \gamma \sigma_0 = [(1.75 - x)/2] \sigma_0. \quad (3)$$

The value of $x$ in Eq. (3) corresponds to the optimal superconducting state, and has been determined by the examination of experimental data in Sec. II to be approximately 0.45. Unfortunately, structural data for $x = 0.45$ are not available, so $T_{C0}$ is calculated for both $x = 0.4$ and $0.5$ using structural data from x-ray diffraction for sintered and oxygenated samples in Ref. 1. For $x = 0.40$, $A = a^2 = (3.876 \text{ Å})^2 = 15.0234$ Å$^2$ and $\zeta = 2.1593$ Å, where $a$ is the basal lattice parameter and $\zeta$ is the difference in $z$ heights between the Ba(La) and O(2)-plane sites.[5] Given $\sigma = \gamma \sigma_0 = [(1.75 - 0.40)/2] 0.228 = 0.1539$, one obtains $\ell = (A/\sigma\eta)^{1/2} = 6.9863$ Å, $(\ell\zeta)^{-1} = 0.06629$ Å$^{-2}$, and Eq. (1) yields $T_{C0} = 82.69$ K. Notice that the calculated σ = 0.1539 is nearly the same as the cuprate-plane holes $n_p = 0.16 \pm 0.02$ for optimum oxygen obtained in Ref. 29. Similarly for $x = 0.50$, $A = (3.873 \text{ Å})^2 = 15.0001$ Å$^2$, $\zeta = 2.1001$ Å, $\sigma = 0.1425$, $\ell = 7.2548$ Å, $(\ell\zeta)^{-1} = 0.06564$ Å$^{-2}$, and $T_{C0} = 81.88$ K. Since $T_C$ is rather independent of $x$ in this regime,[6] one can write a calculated average of $\langle(\ell\zeta)^{-1}\rangle = 0.06597$ Å$^{-2}$ (i.e., $\langle\ell\rangle = 7.1206$ Å and $\langle\zeta\rangle = 2.1297$ Å), corresponding to $\langle T_{C0}\rangle = 82.29$ K, which is adopted for the theoretical result. Given the uncertainties in the Rietveld refinements and the range in $x$, the estimated error in $\langle T_{C0}\rangle$ is ±0.44 K; rounding to one significant decimal digit, $\langle T_{C0}\rangle = 82.3(4)$ K. This result for $T_{C0}$ is plotted at $x = 0.45$ in Fig. 1 and compared to $T_C$ values extracted from resistivity and susceptibility data.[1,2,4]

It is important to note that $T_C$ of $(Ca_xLa_{1-x})(Ba_{1.75-x}La_{0.25+x})Cu_3O_y$ appears to increase slightly under hydrostatic pressure (see inset of Fig. 1).[2] Although the data set is rather limited (data only available for pressures ≤ 0.8 GPa), $dT_C/dP$ decreases rapidly as $x$ approaches $0.4 - 0.5$ from below, indicative of optimization at or near ambient pressure.

Figure 4 shows the experimentally determined values of $T_{C0}$ plotted against the respective calculated values of $(\ell\zeta)^{-1}$ for $(Ca_xLa_{1-x})(Ba_{1.75-x}La_{0.25+x})Cu_3O_y$ (star symbol, for an average $x = 0.45$), compared to other cuprates (open circles), Fe-based pnictides and chalcogenides (open triangles), a ruthenate (filled circle), and an organic (filled triangle). The solid line through the data points, totaling 37 compounds, represents Eq. (1) with $k_B^{-1}\beta = 1247.4$ K-Å$^2$ unchanged from the original publication,[5] successfully and accurately predicting $T_{C0}$ to within ±1.35 K (fitting β anew yields statistically unchanged results: $k_B^{-1}\beta = 1246.0 \pm 3.6$ K-Å$^2$ and same accuracy in $T_{C0}$).[31]

### C. Variation of $T_C$ with $x$

A plausible implication of relatively high residual scattering found at low $x$, particularly for $x = 0.1$, is that $T_C$ is depressed by pair-breaking effects. In the presence of a spin-flip scattering rate $\tau_p^{-1}$, the pair-breaking effect yields a transition temperature $T_C$ that is reduced relative to the optimum $T_{C0}$, and is obtained theoretically from the expression,

$$\ln(T_{C0}/T_C) = \psi(\tfrac{1}{2}) - \psi(\tfrac{1}{2} + \hbar\tau_p^{-1}/4\pi k_B T_C), \quad (4)$$

where ψ is the digamma function (see, e.g., Refs. 11 and 32). In the application of this theory in Ref. 32 to YBa$_2$Cu$_3$O$_{7-\delta}$ modified by atomic substitutions, $\tau_p^{-1}$ was taken to scale with an assumed concentration of magnetic impurities; it is further shown in Ref. 11 that disorder, such as the kind produced by doping YBa$_2$Cu$_3$O$_{7-\delta}$ with Pr on Ba sites, yields a variation in $T_C$ in accordance with Eq. (4).

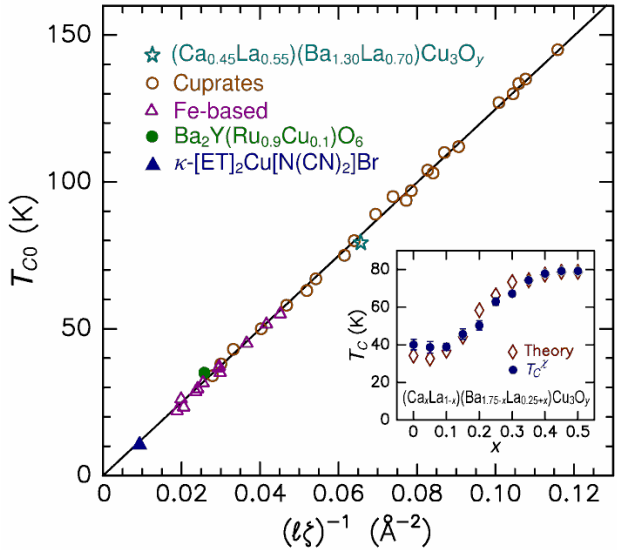

FIG. 4. Experimental $T_{C0}$ vs $(\ell\zeta)^{-1}$, where $\ell$ is intra-layer mean spacing of interacting charges and $\zeta$ is inter-layer interaction distance, for $(Ca_{0.45}La_{0.55})(Ba_{1.30}La_{0.70})Cu_3O_y$ ($x = 0.45$, star symbol), compared to other cuprates (open circles), Fe-based pnictides and chalcogenides (open triangles), a ruthenate (filled circle) and an organic (filled triangle). The solid line through the data points represents Eq. (1). Inset: transition temperatures $T_C^\chi$ of $(Ca_xLa_{1-x})(Ba_{1.75-x}La_{0.25+x})Cu_3O_y$ (filled symbols) and pair-breaking theory [Eq. (4), open symbols] as a function of doping $x$.

Pair-breaking theory is compared with experiment in the inset of Fig. 4. Filled symbols denote data at various $x$ for $T_C^\chi$ determined from magnetic susceptibility (error bars are $\Delta T_C/2$). The open symbols are theoretical $T_C$ from Eq. (4), calculated by modeling the pair-breaking rate as a fraction $\tilde{a}$ of the residual scattering rate and expressed as $\tau_p^{-1} = \tilde{a}\,\tau^{-1}(0)$, where $\tilde{a}$ is an empirical constant and $\tau^{-1}(0)$ is determined from data for the residual resistivity and muon-spin depolarization (interpolation and extension of $x$ dependence was used to determine $\rho(0)$ and $\sigma_\mu$ for which data are unavailable). The fit of Eq. (4) to $T_C^\chi$ yields $T_{C0} = 81.1$ K, $\tilde{a} = 0.24$, and fitting accuracy of 2 K; uncertainty in each parameter is at least 0.5%, as estimated from $\Delta T_C$ and exclusive of other experimental uncertainties (e.g., the systematic error in $\sigma_\mu$). Therefore, the hypothesis of a single optimum parent composition, which becomes non-optimum at reduced $x$ owing to the effects of pair breaking disorder, is consistent with the data on (Ca$_x$La$_{1-x}$)(Ba$_{1.75-x}$La$_{0.25+x}$)Cu$_3$O$_y$. While magnetic impurities have not been reported for the various $x$ at optimum $y$, the spin-flip scattering may be connected to local moment formation in superconducting material with underdoped $y$.[20]

## IV. DISCUSSION

The new datum for (Ca$_{0.45}$La$_{0.55}$)(Ba$_{1.30}$La$_{0.70}$)Cu$_3$O$_y$ included in Fig. 4 provides additional support for the universality of the high-$T_C$ model described herein and elsewhere.[5,12,13] The adherence of this unique charge-compensated compound to Eq. (1) also confirms that inhomogeneity and magnetism are incompatible with optimal high-$T_C$ superconductivity. Based on interlayer Coulomb coupling, and recognizing that measurements made only on optimal materials will necessarily reflect the intrinsic superconducting state, this theory has been shown to be valid for 37 different optimal compounds representing six superconducting families. As described in Sec. III, the interacting charge density $\ell^{-2}$ follows from the charge fraction $\sigma$ that, in itself, is not a universal constant among the various optimal cuprate compounds [Eq. (2)]. This stands in stark contrast with earlier models representing $T_C$ as a function solely of a doping parameter,[33,34] which do not distinguish the carriers actually participating in the pairing interaction. Such models were recently applied to (Ca$_x$La$_{1-x}$)(Ba$_{1.75-x}$La$_{0.25+x}$)Cu$_3$O$_y$ for the purposes of XAS calibration[29] and for asserting inconsistency.[35] Another significant deficiency in models restricted to doping dependence (e.g., as in Ref. 36 for YBa$_2$Cu$_3$O$_{7-\delta}$) is the omission of the interaction distance $\zeta$; inclusion of this structural length in Eq. (1) proves to be essential for universal accuracy in predicting $T_C$, given that high-$T_C$ superconductivity originates from interlayer Coulomb interactions.

While the negligible XAS spectral response with $x$ observed for the planar Cu and O ions[29] is seemingly in conflict with transport and susceptibility measurements showing a significant correlation of $T_C$ with $x$ (Fig. 1), and $\mu^+$SR experiments finding significant variation of $\sigma_\mu$ with $x$ (e.g., the two square data points of Fig. 3 and in Ref. 6), it does indicate that the variation of $T_C$ with cation substitution[6,20-22,35] is unrelated to the two-dimensional (2D) carrier density in the cuprate planes.[37] Implicit in this dichotomy is that $\mu^+$SR measurements (and other techniques) are insensitive in probing superconductivity associated with the CuO$_2$ planes. Taken together, these studies therefore locate the superconducting hole condensate in the type I reservoirs (La-substituted BaO-CuO-BaO structures). Experimental uncertainty in XAS allows only for a possible 12% of the superconducting fraction to reside in the cuprate layers.

In much of the previously published work on Ca$_x$La$_{1-x}$(Ba$_{1.75-x}$La$_{0.25+x}$)Cu$_3$O$_y$ it was assumed that the doping (both $x$ and $y$) can be varied rather widely without affecting the quality of the superconducting condensate. Each value of $x$ was assumed to constitute a different cuprate family, owing to variations in structure (lattice parameters and oxygen buckling angles)[21] and critical doping levels.[22] Such a designation assumes equal significance among the local maxima in $T_C$ associated with each $x$, which is incommensurate with the experimental evidence for degradation of the superconducting condensate for $x < 0.4$ (for any given value of $y$), exhibited, e.g., by the corresponding resistivities, susceptibilities, transition widths, and residual carrier scattering. From Figs. 1 and 2, it is evident that there exists only one unique optimal stoichiometry, which is achieved for $x$ between 0.4 and 0.5.

The initial offset introduced by the substitution of 12.5% La$^{+3}$ into the Ba$^{+2}$O layers of the parent compound $LaBa_2Cu_3O_{7-\delta}$ reduces the charge available for superconductivity, thereby limiting the ultimate value of $T_{C0}$ achievable. The initial imbalance of charge introduced between the type I and type II reservoirs is clearly observed from the pressure dependence (see inset of Fig. 1), which shows a decreasing $dT_C/dP$ as $x$ approaches 0.4 from below. Application of hydrostatic pressure facilitates charge transfer along the hard axis; the structurally related compound, $YBa_2Cu_4O_8$, is a good example of the effect of pressure in optimizing $T_C$.[38] The fact that $dT_C/dP$ approaches zero for $x \sim 0.4$ indicates an ambient-pressure charge equilibrium corresponding to the one optimal stoichiometry.

### A. Trends in optimal oxygen doping

Obtaining the charge fraction $\sigma$ by scaling to $\sigma_0$ of the optimal compound $YBa_2Cu_3O_{6.92}$ is predicated on implicit optimization of the oxygen stoichiometry in the high-$T_C$ cuprates under consideration. Figure 5 examines this by plotting the oxygen content $y$ against $T_C$ for two parent compounds of $(Ca_xLa_{1-x})(Ba_{1.75-x}La_{0.25+x})Cu_3O_y$, namely $LaBa_2Cu_3O_y$ [$T_{C0}$ = 97 K from Ref. 39 ($y$ not given), $y$ = 6.85 from Ref. 40, filled square[41]] and $YBa_2Cu_3O_y$ ($T_{C0}$ = 93.78 K, $y$ = 6.92 from Ref. 42, filled triangle), corresponding to optimum $y$ for each, and for $(Ca_xLa_{1-x})(Ba_{1.75-x}La_{0.25+x})Cu_3O_y$, where $y$ corresponds to highest (resistive onset) $T_C$ values of 57.7, 68.8, 76.6, and 81.5 K for $x$ = 0.1, 0.2, 0.3, and 0.4, respectively (open circles).[4] The filled circle corresponds to $x$ = 0.45, $y$ = 7.135 and $T_{C0}$ = 80.5 K, as determined for optimal $(Ca_xLa_{1-x})(Ba_{1.75-x}La_{0.25+x})Cu_3O_y$ in Sec. III B from data in Ref. 1. A dashed line is fitted to the data for the three optimum compounds (filled symbols), indicating the linear trend [slope $-0.017(9)$ K$^{-1}$, intercept 8.50(8), linear correlation coefficient $R^2$ = 0.997, and fitting error $\delta y$ = 0.011]. Three open circle symbols corresponding to $(Ca_xLa_{1-x})(Ba_{1.75-x}La_{0.25+x})Cu_3O_y$ with non-optimum doping ($x < 0.4$) deviate to the left of this trend line. Of particular interest is that the 40% variation in $T_C$ with $x$ is associated with rather small changes in $y$, i.e., the trend in $(Ca_xLa_{1-x})(Ba_{1.75-x}La_{0.25+x})Cu_3O_y$ is very much weaker than expected for a series of optimum compounds (dashed line). Near invariance of $y$ suggests that the charge balance between the two reservoir types changes with the charge compensation variable $x$, which is non-optimum for $x < 0.4$, where depressed and broadened transitions are observed (Figs. 1 and 2).

Combining the various reported results for optimal superconductivity in $(Ca_xLa_{1-x})(Ba_{1.75-x}La_{0.25+x})Cu_3O_y$, the optimal oxygen content is estimated as $y_0$ = 7.15 ± 0.02, the optimal charge compensation variable is in the range $0.4 \leq x \leq 0.5$, and the unique optimal $T_{C0}$, determined by various methodologies, is in the range 80.5 to 82 K.[1-4,21,29,35]

For the oxygen underdoped compositions, the onset of superconductivity occurs in the vicinity of a threshold $y_c$ that systematically decreases with increasing $x$. According to Fig. 1 of Ref. 20, $y_c \approx 6.97$ at $x$ = 0.1 and $y_c \approx 6.88$ at $x$ = 0.4. Since superconductivity is already degraded at $x$ = 0.1 (e.g., insets in Figs. 2 and 4), it is logical that less relative oxygen doping ($y_0 - y_c$) = 0.17 (compared to 0.27 at $x$ = 0.4) is sufficient for destroying superconductivity. However, one recognizes that alternative interpretations of the $x$-$y$ behavior have proposed various $x$-dependent factors to scale the $y$ dependence.[2,6,20-22,35] In one of these, comparing $^{17}O$ NQR data at $x$ = 0.1 and 0.4, variations in superconductivity and magnetism are attributed to an $x$ dependence in the efficiency of hole injection.[22] Remaining unaccounted for by various $x$-dependent scaling methods are the $O_{1s}$ $K$-edge XAS measurements that find no $x$ dependence in the carrier concentration at the cuprate-plane O site.[29] A plausible explanation is that degradation of the Ca/La and Ba/La stoichiometries tends to be compensated by adjustments in ($y_0 - y_c$). Experimental results, particularly the near invariance of $y_0$, therefore validate the unique value for $T_{C0}$, which is used in the analysis of $T_C$ according to Eq. (4).

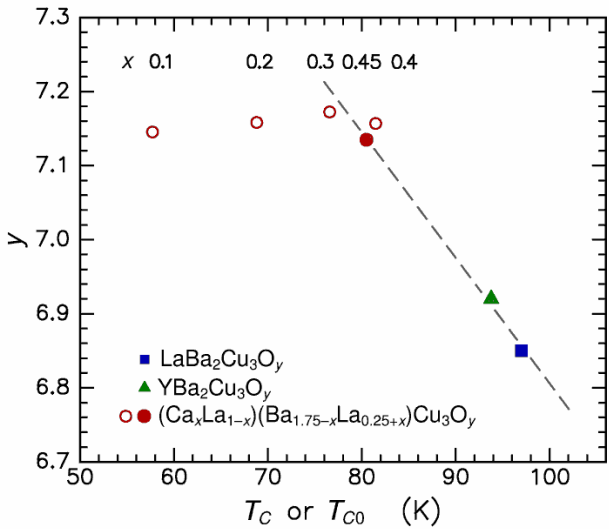

FIG. 5. Variation of oxygen parameters $y$ with transition temperatures for $LaBa_2Cu_3O_y$ and $YBa_2Cu_3O_y$ (filled symbols, $T_{C0}$ at optimum $y$), and for $(Ca_xLa_{1-x})(Ba_{1.75-x}La_{0.25+x})Cu_3O_y$ from resistive onset $T_C$ at $x$ = 0.1, 0.2, 0.3, and 0.4 after Ref. 4, open circles, and experimental optimum $T_{C0}$ at $x$ = 0.45 (see Sec. III B), filled circle. Dashed line denotes linear fit to the optimum compounds.

## B. Inhomogeneous superconductors

Superconducting transition temperatures measured as a function of doping typically exhibit a "superconducting dome" effect (a rise in $T_C$ followed by a fall, as stoichiometry passes through the underdoped-optimal-overdoped regimes); this same behavior was found in $(Ca_xLa_{1-x})(Ba_{1.75-x}La_{0.25+x})Cu_3O_y$ by the non-monotonic variation of $T_C$ with $y$ at constant $x$.[4,21] Figure 1 illustrates that $T_C$ also arches over to a maximum as a function of $x$ (at optimum $y$), accompanied by a sharpening of the superconducting transition; the region $x < 0.4$ therefore exhibits the hall marks of percolative inhomogeneous superconductivity in the underdoped regime (see, e.g., Ref. 43; formulations with $x \geq 0.6$ turn out to yield phase-separated samples[1]).

The $\mu^+$SR data of Fig. 3 suggest a similar tendency for $T_C$ to approach a maximum as a function of $\sigma_\mu/\gamma_\mu$. For underdoped cuprates the variation of $T_C$ with $\sigma_\mu$ is of interest in connection with the pseudogap phenomenon, identified by a characteristic temperature $T^* > T_C$.[19,43-45] For example, a resonating valence bond model applied to calculating the penetration depth has found that the "dome" characteristic of $T_C$ vs doping translates into a similar non-linear dependence of $T_C$ on $\lambda^{-2}(0)$ that has been favorably scaled to experimental data on selected compounds.[45]

The dashed line of Fig. 3 has been considered in terms of a granular texture model of inhomogeneous superconductivity from which a similar proportionality between $T_C$ and $\lambda^{-2}(0)$ has been derived theoretically (coefficient reproduced to within a factor of 2).[46] In a compendium of experimental data that includes nonoptimum compounds, it was found that $T_C$ multiplied by the normal-state conductivity is proportional to $\lambda^{-2}(0)$ with about ± 20% error; this correlation is attributed to the influence of disorder scattering.[47] Universal proportionality between $T_C$ and $\sigma_\mu$, or the amended nearly linear scaling discussed in Ref. 27, are at best qualitative concepts[19] that were shown to be in disagreement with $\mu^+$SR data.[48] Moreover, the relation $\sigma_\mu \propto n_S/m^*$ can only be considered an idealization, uncorrected for perturbations by materials inhomogeneity, electronic disorder and fluxon pinning effects (see, e.g., Ref. 16). This is clearly demonstrated in Fig. 3 in the case of $(Ca_xLa_{1-x})(Ba_{1.75-x}La_{0.25+x})Cu_3O_y$,[49] and also by the data for $YBa_2Cu_3O_{7-\delta}$,[23] where $T_C$ for samples near the optimum 90-K phase varies by 1.2% (91.3 to 92.4 K), yet results for $\sigma_\mu$ vary by 40 %, corresponding to a factor of 1.5, with most of these data falling to the right of the (extrapolated) dashed line of Fig. 3.

Specific evidence that $x = 0.1$ tends to form inhomogeneous material is found in resistivity and magnetization data on $(Ca_xLa_{1-x})(Ba_{1.75-x}La_{0.25+x})Cu_3O_y$ single crystals grown with optimum $y$.[50] In Fig. 3 of Ref. 50, an $x = 0.1$ crystal exhibits a very broad transition, extending from an onset near 79 K to full superconductivity below 50 K, clearly reflecting an inhomogeneous superconducting state. A similarly prepared crystal with $x = 0.4$ exhibits sharp resistive and magnetic transitions with an onset near 81 K, consistent with results for ceramic specimens. Another anomaly associated with $x = 0.1$ is the absence of a metallic transport at overdoping in $y$ where $T_C = 0$, as revealed by a minimum in the temperature dependence of $\rho(T)$.[51] Interpreted as absence of a quantum critical point in $y$,[51] this finding makes it difficult to treat formulations with $x = 0.1$ as a separate family of $(Ca_xLa_{1-x})(Ba_{1.75-x}La_{0.25+x})Cu_3O_y$ compounds. The broadened transitions also create difficulties in quantifying a superconducting dome ($T_C$ vs $y$); for example, at $x = 0.1$ and $y = 6.986$, which is underdoped and above the minimum for superconductivity, the claimed $T_C$ value of about 16 K entered in the phase diagram corresponds to a specimen that is non-superconducting above 5 K and has a negative resistivity slope between 25 and 80 K (see Fig. 6 of Ref. 4).[4]

## C. Superconductivity and magnetism in $(Ca_xLa_{1-x})(Ba_{1.75-x}La_{0.25+x})Cu_3O_y$

In a series of presentations of experimental studies, the superconductivity was related to magnetism in underdoped $(Ca_xLa_{1-x})(Ba_{1.75-x}La_{0.25+x})Cu_3O_y$ through various constructions of reduced phase diagrams, where changes in $y$ were scaled by factors varying with $x$.[20,22,35,52] The differing approaches (e.g., Refs. 20 and 22) show the commonality in how magnetism becomes disordered and diminishes, concomitant with the emergence of superconductivity, as $y$ is increased from 6.4 to 7.15 (a difference of 0.75). Part of that commonality is exhibited by the variation of Néel temperature $T_N$ with $x$ and $y$, where an $x$-independent crossing (or stationary) point occurs at $y \sim 6.5$ and $T_N \approx 380$ K. Although $T_N$ is defined by a sharp phase transition at extreme oxygen depletion ($y \sim 6.4$), significant broadening becomes evident for $y$ near the spin-glass region (e.g., $T_N = 221$ K of width > 35 K for $x = 0.1$, $y = 6.858$, and where $T_C = 0$),[20] indicating that phase diagram analysis encompasses materials with inhomogeneous magnetism. Evidently, the values $y_N$ constructed in Ref. 20 from $y$ dependence in $T_N$ connote the onsets of inhomogeneity in antiferromagnetic ordering, which is consistent with the generally inhomogeneous spin-glass freezing.[20,52] Since phase separation is not expected to broaden $T_N$, critical oxygen levels such as $y_N$ point to intrinsic magnetic inhomogeneity,[52] rather than a possibility of phase transitions,[22] analogous to how $y_c$ is related to the destruction of superconductivity in the presence of magnetism.

The $\mu^+$SR measurements of superconducting and spin-glass properties of $y$-underdoped $(Ca_xLa_{1-x})(Ba_{1.75-x}La_{0.25+x})Cu_3O_y$, particularly in the region where the two phases coexist, had earlier prompted the suggestion of microscopic phase separation.[52] In a later study using $^{63}Cu$ NQR, a possibility of mesoscopic inhomogeneities remained.[35] However, a more recent work, which included comparisons of $^{17}O$ NQR at $x = 0.1$ and $0.4$, has posited changes in oxygen doping efficiency.[22] Generally shown in these studies is that underdoping in $x$ allows for less underdoping in $y$, i.e., underdoping in both $x$ and $y$ depletes carriers synergistically. The most notable feature common to all of the various scaled phase diagrams constructed from these observations[20,22,35,52] is the nearly $x$-independent oxygen doping at highest $T_C$, which is consistent with a unique optimal superconducting composition, and in contrast to the shifts observed in the magnetic transitions.

The authors of Ref. 20 characterized their results as lending strong support for magnon-mediated pairing in the cuprates. Subsequent work has reported absence of an oxygen isotope effect (OIE) on $T_N$ in $(Ca_xLa_{1-x})(Ba_{1.75-x}La_{0.25+x})Cu_3O_y$ with oxygen content $y < 6.6$ (where $T_N$ is independent of $y$),[53] indicating that the magnetic ordering is independent of isotopic substitution. Together with the presumed OIE on $T_C$ (e.g., the comparison with Pr-doped $YBa_2Cu_3O_{7-\delta}$ in Ref. 53), this finding clearly disconnects magnetism from superconductivity in $(Ca_xLa_{1-x})(Ba_{1.75-x}La_{0.25+x})Cu_3O_y$.

Implicit in the behavior observed is that magnetism accompanies sample degradation and suppression of superconductivity. It therefore seems reasonable to attribute these effects to increased pair breaking, as defined by Eq. (4), in the underdoped regime, exclusive of any interdependency or symbiotic relationship between magnetism and superconductivity in $(Ca_xLa_{1-x})(Ba_{1.75-x}La_{0.25+x})Cu_3O_y$. Although these magnetic studies are useful from the perspective of understanding the nonoptimal compounds, they shed limited light on the properties of the optimized superconducting condensate, which is the focus of the present work.

## V. CONCLUSIONS

Given the body of evidence from transition-width, resistivity, and $\mu^+$SR data (Figs. 2 and 3), it is clear that there exists only one optimal $(Ca_xLa_{1-x})(Ba_{1.75-x}La_{0.25+x})Cu_3O_y$ stoichiometry; $x \approx 0.45 \pm 0.05$ and $y \approx 7.15 \pm 0.02$. This stoichiometry also corresponds to a precipitous decrease in $dT_C/dP$, indicating that optimization occurs at ambient pressure. By applying rule (2b), the fractional charge $\sigma$ for optimal $(Ca_xLa_{1-x})(Ba_{1.75-x}La_{0.25+x})Cu_3O_y$ (scaled to $\sigma_0$ for $YBa_2Cu_3O_{6.92}$) is obtained as $\sigma = [(1.75 - x)/2]\sigma_0$, which is the charge available for superconductive pairing between the nearest-neighbor Ba/LaO and $CuO_2$ layers (separated by an interaction distance $\zeta$) and evaluates to $\sigma = 0.1482 \pm 0.0057$, including the uncertainty in the optimal $x$. Given the relative insensitivity of $T_C$ to $x$ in the range $0.40 \leq x \leq 0.50$ (with optimized $y$), averaging $(\ell\zeta)^{-1}$ for the two end points provides an average value for the optimal transition temperature of $T_{C0} = 82.3(4)$ K, which is in excellent agreement with the experimental value of 80.5 K presented in Figs. 1, 4 and 5, as well as the claimed $T_C^{max}$ of 82 K in Ref. 2 and 81 K in Refs. 3 and 4.

The local maxima in transition temperature observed for $x < 0.4$ (tuned by adjusting $y$), while not reflective of optimal superconducting states (independently indicated in Fig. 5), do nevertheless provide insights into sample inhomogeneities and magnetic phenomenology in the underdoped region, specifically showing that (1) $T_C$ is not proportional to the $\mu^+$SR depolarization rate $\sigma_\mu$ (Fig. 3) and (2) the magnetism observed, at least for severely oxygen-deficient samples, is unrelated to the superconductivity. Moreover, the pair-breaking formalism of Eq. (4) with a unique and intrinsic optimum $T_{C0}$ is consistent with the observed variation of $T_C$ with $x$.

Particularly enlightening information comes from XAS experiments,[29] which find a near-invariance of the charge per planar Cu and O (corresponding to a near-constant 2D carrier density in the cuprate planes) as a function of $x$. Given that the $\mu^+$SR experiments conducted on $(Ca_xLa_{1-x})(Ba_{1.75-x}La_{0.25+x})Cu_3O_y$ record a significant dependence between $\sigma_\mu$ and $x$ (consistent with the $x$ dependence of $T_C$ from transport and susceptibility),[6] the results of Ref. 29 imply that most (e.g., 88%) of the superconducting condensate measured by $\mu^+$SR (and other techniques) is external to the $CuO_2$ planes. This leaves the (Ba/La)O layers of the type I reservoir as the likely locus of the superconducting hole condensate, which is an assignment consistent with that considered for $YBa_2Cu_3O_{7-\delta}$, as discussed in Ref. 5.

With the addition of $(Ca_{0.45}La_{0.55})(Ba_{1.30}La_{0.70})Cu_3O_{7.15}$, the list of compounds which behave in a manner consistent with the theoretical expression of Eq. (1) has grown to 37 in number, from six different superconducting families.

## ACKNOWLEDGMENTS

We are grateful for the support of the Physikon Research Corporation (Project No. PL-206) and the New Jersey Institute of Technology. We also thank Amit Keren for suggesting this problem. This work has been published in Physical Review B.[55]